\documentclass[
reprint,
aps,
twocolumn,
superscriptaddress, 
floatfix,
]{revtex4}

\usepackage{physics}

\usepackage{gensymb}
\usepackage{amssymb}
\usepackage{bm, amsmath}
\usepackage{bm}% bold math
\usepackage{physics}
\usepackage[euler]{textgreek}
\usepackage{siunitx}
\usepackage[hidelinks]{hyperref}
\usepackage[version=4]{mhchem}
\usepackage{natbib}
\usepackage{balance}

\usepackage{graphicx}% Include figure files
\usepackage[caption=false]{subfig}
\usepackage[outdir=./]{epstopdf}

\usepackage[normalem]{ulem} %strikeout command: \sout
\usepackage{verbatim}
\usepackage[draft]{todonotes}
\usepackage{float}

\newcommand{\angstrom}{\textup{\AA}}

\begin{document}
\title{Giant anomalous Hall effect in epitaxial Mn$_{3.2}$Ge films with a cubic kagome structure}

\author{J.~S.~R.~McCoombs}
\affiliation{Department of Physics and Atmospheric Science, Dalhousie University, Halifax, Nova Scotia, Canada B3H 3J5}

\author{B.~D.~MacNeil}
\affiliation{Department of Physics and Atmospheric Science, Dalhousie University, Halifax, Nova Scotia, Canada B3H 3J5}

\author{V.~Askarpour}
\affiliation{Department of Physics and Atmospheric Science, Dalhousie University, Halifax, Nova Scotia, Canada B3H 3J5}

\author{J.~Myra}
\affiliation{Department of Physics and Atmospheric Science, Dalhousie University, Halifax, Nova Scotia, Canada B3H 3J5}

\author{H.~Herdin}
\affiliation{Department of Physics, Simon Fraser University, Burnaby, British Columbia, Canada, V5A 1S6}

\author{M.~Pula}
\affiliation{Department of Physics and Astronomy, McMaster University, Hamilton, Ontario, Canada L8P 4N3}

\author{M.~D.~Robertson}
\affiliation{Department of Physics, Acadia University, Wolfville, Nova Scotia, Canada B4P 2R6}

\author{G.~M.~Luke}
\affiliation{Department of Physics and Astronomy, McMaster University, Hamilton, Ontario, Canada L8P 4N3}

\author{K.~L.~Kavanagh}
\affiliation{Department of Physics, Simon Fraser University, Burnaby, British Columbia, Canada, V5A 1S6}

\author{J.~Maassen}
\affiliation{Department of Physics and Atmospheric Science, Dalhousie University, Halifax, Nova Scotia, Canada B3H 3J5}

\author{T.~L.~Monchesky}\thanks{tmonches@dal.ca}
\affiliation{Department of Physics and Atmospheric Science, Dalhousie University, Halifax, Nova Scotia, Canada B3H 3J5}
\date{\today}

\begin{abstract}
We report on the first example of epitaxial Mn$_{3 + \delta}$Ge thin films with a cubic $L1_2$ structure. The films are found to exhibit frustrated ferromagnetism with an average magnetization corresponding to 0.98~$\pm$~0.06~$\mu_B$/Mn, far larger than the parasitic ferromagnetism in hexagonal Mn$_3$Ge and the partially compensated ferrimagnetism in tetragonal Mn$_3$Ge. The Hall conductivity is the largest reported for the kagome magnets with a low temperature value of $\sigma_{xy} = 1587$~S/cm.  Density functional calculations predict that a chiral antiferromagnetic structure is lower in energy than a ferromagnetic configuration in an ordered stoichiometric crystal. However, chemical disorder driven by the excess Mn in our films explains why a frustrated 120$\degree$ spin structure is not observed. Comparisons between the magnetization and the Hall resistivity indicate that a non-coplanar spin structure contributes the Hall signal. Anisotropic magnetoresistance and planar Hall effect with hysteresis up to 14~T provides further insights into this material. 
\end{abstract}

\maketitle

\section{Introduction}
\begin{figure*}[htbp]
    \centering
	\includegraphics[width=2.0\columnwidth]{{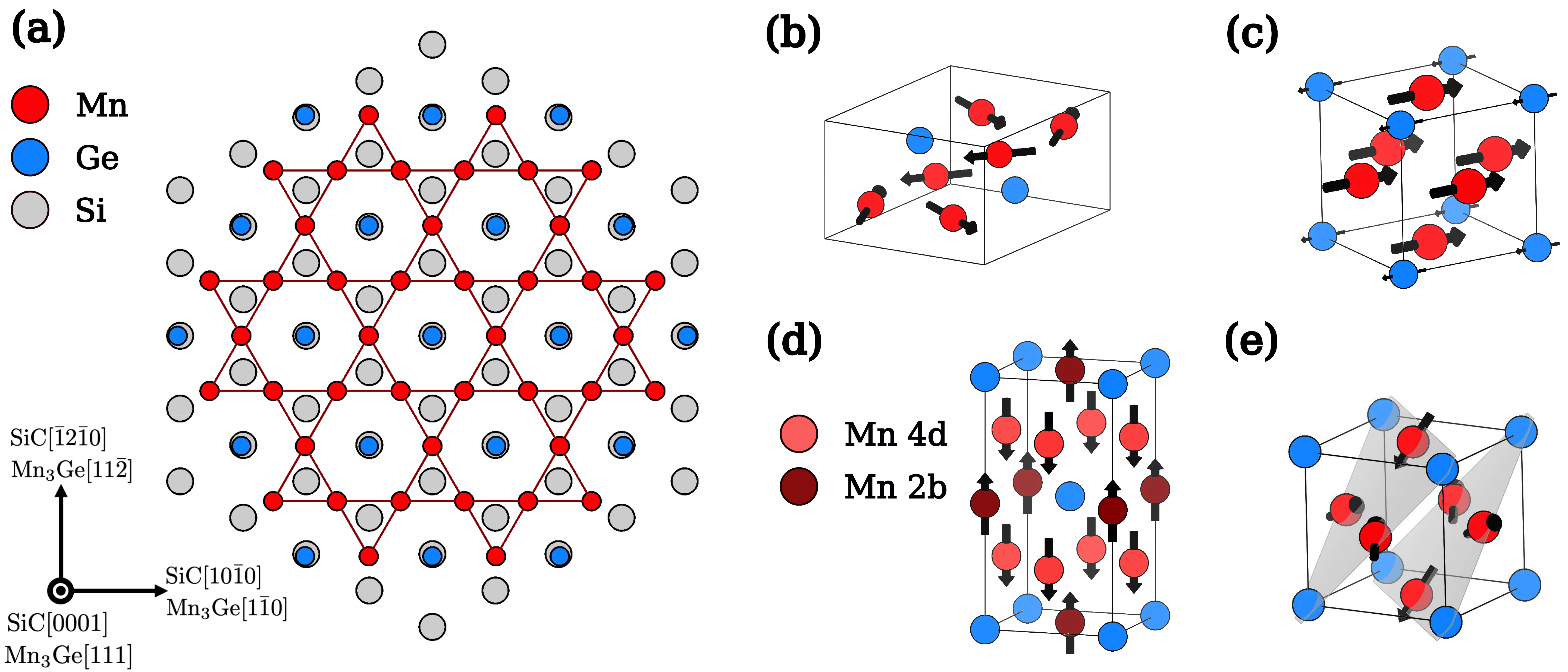}}
	\caption{(a) The expected epitaxial relationship of $L1_2$ Mn$_3$Ge atop SiC. (b) The crystal structure and inverse-triangular, AF spin-configuration of hexagonal $D0_{19}$ Mn$_3$Ge. (c) The crystal structure and DFT calculated ferromagnetic spin-configuration of cubic $L1_2$ Mn$_3$Ge. (d) The crystal structure and ferrimagnetic spin-configuration of $D0_{22}$ Mn$_3$Ge. (e) The cubic $L1_2$ structure with DFT calculated triangular, AF spin-configuration. Images generated with \textit{VESTA}~\cite{Momma2008}.}
	\label{fig:Structures}
\end{figure*}

Magnetic topological metals offer opportunities to explore the entanglement of real-space and momentum space spin-structures.  In Weyl semimetals formed by broken time reversal symmetry, the real-space spin texture provides a way of manipulating the topological electronic states.  In the case of the hexagonal Mn$_3$Sn and Mn$_3$Ge kagome materials, a rotation of the chiral antiferromagnetic spin configuration (shown in Fig.~\ref{fig:Structures}(b)) with an external magnetic field causes a rotation of the Weyl nodes~\cite{Kuroda:2017nm}. The momentum-space Berry curvature in these structures leads to a large anomalous Hall effect (AHE)~\cite{Nakatsuji:2015nat,Kiyohara:2016pra} and anomalous Nernst effect (ANE)~\cite{Ikhlas:2017np,Hong:2020prm} that have created opportunities for developing devices.  The antiferromagnetic chiral structure of Mn$_3$Ge has also been used to manipulate spin-triplet Cooper pairs in a Josephson junction, with potential applications in superconductor logic circuits~\cite{Jeon:2023nn}.

The closely related set of cubic kagome antiferromagnets Mn$_3$X, X = \{Ir, Pt, Rh\} have a positive vector chirality~\cite{Tomeno:1999jap, Yasui:1992jmmm, Kren:1968pr} (see Fig.~\ref{fig:Structures}(e)), opposite to that of Mn$_3$Sn~\cite{Tomiyoshi:1982jpsj} and Mn$_3$Ge~\cite{Tomiyoshi:1983jmmm}.  Their topological band structures give rise to large anomalous Hall~\cite{Liu:2018ne, Zuniga-Cespedes:2023njp} and spin-Hall effects~\cite{Zhang:2016sa}.  The large uniaxial anisotropy in the case of Mn$_3$Ir leads to a canting of the spins out of the kagome planes and to a scalar spin chirality $\vb S_i \cdot (\vb S_j \times \vb S_k)$.  However, it is spin-orbit coupling and broken mirror symmetry of the spin structure that gives rise to large momentum-space Berry curvature in this material~\cite{Chen:2014prl}.

In this paper we explore the properties of Mn$_{3.2}$Ge films, which, in fact, has very similar hexagonal, cubic and tetragonal polytypes with dramatically different magnetic properties.  The phase that is obtained depends sensitively on the substrate and growth conditions. In bulk crystals, the hexagonal $\epsilon$-Mn$_{3}$Ge phase ($D0_{19}$ crystal structure with space group \#194, $P6_3/mmc$), is obtained by annealing at temperatures above 500~$\degree$C~\cite{Ohoyama:1961jpsj, Yamada:1988pb}. The lattice parameters are $a_h=0.534$~nm and $c_h = 0.431$~nm~\cite{Ohoyama:1961jpsj}. The Mn atoms occupy the $6h$ Wyckoff sites and form an AB-stacking of kagome layers with their spins pointing in the plane at an angle of 120$\degree$ to their neighbors~\cite{Kadar:1971ijm}. This order exists below a Curie temperature of 365~K~\cite{Yamada:1988pb}. The in-plane easy axis magnetocrystalline anisotropy induces a weak moment of 0.007~$\mu_B$/Mn~\cite{Yamada:1988pb} in the kagome plane~\cite{Tomiyoshi:1983jmmm}. Epitaxial Mn$_{3.22}$Ge(0001) films were achieved on sapphire (0001) substrates with, a Ru buffer layer, by co-sputtering Mn and Ge at a rate of 10~nm/min and subsequent annealing at 500~$\degree$C~\cite{Hong:2020prm}, as well as by molecular beam epitaxy on LaAlO$_3$(111) at a temperature of $570\degree$C~\cite{Hong:2022aplm}.

Tetragonal $\epsilon_1$-Mn$_{3}$Ge ($D0_{22}$ crystal structure with space group \#139, $I4/mmm$ shown in Fig.~\ref{fig:Structures}(d)) is the thermodynamically stable phase below 500$\degree$C, which possesses ferrimagnetic order with spins parallel to the c-axis~\cite{Ohoyama:1961jpsj} and an estimated Curie temperature of 650$\degree$C~\cite{Kadar:1971ijm}.  This phase, with lattice parameters $a_{t} = 0.3816$~nm and $c_{t} = 0.7216$~nm~\cite{Kadar:1971ijm}, grows epitaxially on SrTiO$_3$(001), which has a lattice parameter 2.3\% larger than the a-parameter~\cite{Kurt:2012apl}. Epitaxial films have also been grown on MgO(001) with a Cr buffer layer~\cite{Sugihara:2014apl}. The films have strongly compensated moments with a net magnetization $M_s =73$~kA/m, and a very high uniaxial anisotropy $K_u =0.91$~MJ/m$^3$ despite the absence of heavy elements.  The tetragonal phase can be transformed into the cubic variant by stretching the unit cell by 5\%  along the c-axis and interchanging one of the Ge atoms on the $2a$-sites with one of the Mn atoms on the $2b$-sites.  In this way, the tetragonal phase can be viewed as a distorted ABC-stacked kagome structure.

Cubic Mn$_{3}$Ge ($L1_2$ structure with space group \#221, $Pm\bar{3}m$) differs from the hexagonal phase only in its ABC-stacking of the kagome layers.  The Mn atoms sit on the $3c$ Wyckoff sites, corresponding to the cubic unit cell faces. There is only one report to date on the synthesis of this material, which was achieved by reacting Mn and Ge at high temperatures and pressures~\cite{Takizawa:2002jpcm}. The cubic lattice parameter $a=0.3802(1)$~nm, is close to both the $a_t$ and $a_h/\sqrt{2}$ parameters. However, the magnetic properties are quite different: magnetometry measurements revealed ferromagnetic behavior, with a saturation magnetization that corresponds to $0.87~\mu_B$/Mn and a Curie temperature, $T_C$, of approximately 400~K. Here, we report the first demonstration of the growth of cubic Mn$_{3 + \delta}$Ge films.  We used 4H-SiC(0001) substrates, which have a $-0.75\%$ lattice mismatch.

The rest of this paper is structured as follows: We first describe the growth procedure and outline the structural experiments we conducted which lead us to the conclusion that we have an epitaxial, cubic kagome system. We probe the spin-configuration in this system with complementary in-plane and out-of-plane magnetometry and DC transport measurements. Finally, we present a density functional theory (DFT) investigation of the ground state spin structure of stoichiometric Mn$_3$Ge and the resulting Berry curvature.
\section{Growth}
We used 4H-SiC(0001)$\pm 0.5\degree$ Si-terminated wafers with a resistivity of $\sim 1\times10^{5}~\Omega$cm as substrates. Prior to growth, these substrates were degreased in acetone and methanol, rinsed in deionized water, dried with nitrogen, then immediately loaded into a VG--V80 SiGe molecular beam epitaxy (MBE) chamber with a base pressure of less than $4\times10^{-11}$~Torr. The substrates were then degassed at $600\degree$C for 12 hr, followed by flashing to $1100\degree$C for 1 hr to remove the SiO$_2$ layer. This process results in a $\left(\sqrt{3}\times\sqrt{3}\right)R30\degree$ surface reconstruction free of any oxygen, as is evidenced by \textit{in-situ} reflection high-energy electron diffraction (RHEED) and Auger electron spectroscopy (AES). The substrate is then cooled at a rate of approximately $20\degree$C/min to room temperature, $T_R$, for co-deposition of Mn via effusion cell and Ge via electron beam evaporation. Upon reaching $T_R$, Mn and Ge are deposited at an atomic ratio of 5:1. The Mn rate is monitored with a calibrated ion gauge, while the Ge rate is monitored with a quartz oscillator. Upon completion of co-deposition, the sample is heated to $600\degree$C for 1 hr while monitored with RHEED until diffraction spots are recovered. The temperature of $600\degree$C was chosen since it was found that lower temperatures led to smoother films. However, temperatures up to $\sim550\degree$C did not crystallize. Thus, $600\degree$C seems to be near the lowest temperature which results in a crystallized film.
%
%\begin{figure}[h]
%    \centering
%	\includegraphics[width=1.0\columnwidth]{{RHEED.png}}
%	\caption{RHEED images along the two principle in-plane directions showing (a,b) the $\left(\sqrt{3}\times\sqrt{3}\right)R30\degree$ reconstruction of the clean SiC(0001) surface and (c,d) a $(1\times1)$ reconstruction of the Mn$_3$Ge film after annealing has taken place.}
%	\label{fig:RHEED}
%\end{figure}
%

Due to Mn evaporation from the film at elevated substrate temperatures, a ratio Mn:Ge = 5:1 was required to achieve the desired stoichiometery, as is evidenced by energy dispersive X-ray spectroscopy (EDS). For this reason, we use the total deposited thickness of Ge as a calibration of the expected thickness of our Mn$_3$Ge films. The number density of elemental Ge is $n_{\text{Ge}}^{\text{Ge}} = 0.04412~\angstrom^{-3}$, while the number density of Ge in Mn$_3$Ge is $n_{\text{Mn}_3\text{Ge}}^{\text{Ge}} = 0.0182~\angstrom^{-3}$. Hence, for a thickness of $t_{\text{Ge}}$ deposited we expect a thickness of
\begin{equation}
    t_{\text{Mn}_3\text{Ge}} = \frac{n_{\text{Ge}}^{\text{Ge}}}{n_{\text{Mn}_3\text{Ge}}^{\text{Ge}}}\frac{t_{\text{Ge}}}{\alpha}
    \label{eq:Thickness}
\end{equation}
for our film, where $\alpha=0.8$ accounts for the remaining excess Mn occupying Ge sites. Lastly, before removing the sample from the MBE, it is again cooled to $T_R$ for the deposition of a 10~nm protective, amorphous layer of Si.

\section{Structural and chemical characterization}

\subsection{Diffraction}
\begin{figure}[htb]
    \centering
	\includegraphics[width=1.0\columnwidth]{{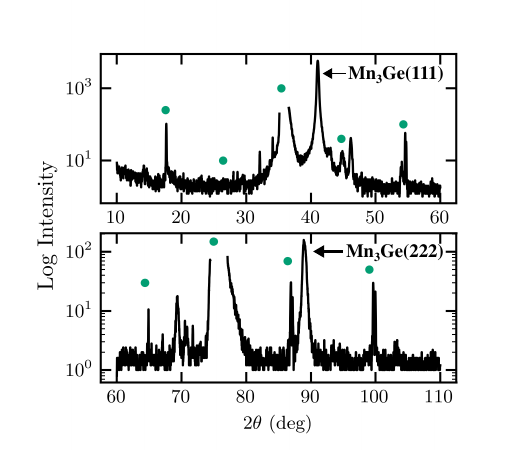}}
	\caption{Log-scale plot of traditional $\theta$-$2\theta$ diffraction pattern using Cu-K$\alpha$ X-rays. The SiC(0004) and SiC(0008) reflections are omitted. Green circles indicate positions of allowed and forbidden SiC(000$\ell$) reflections.}
	\label{fig:TH_TTH_Plot}
\end{figure}

After aligning the SiC(0004) reflection along the $q_z$ axis of the Siemens D500 X-ray diffractometer, traditional $\theta$-$2\theta$ measurements are performed using a Cu source and monochrometer which allows only Cu K-$\alpha$ wavelengths. Figure~\ref{fig:TH_TTH_Plot} shows the diffraction pattern obtained exhibiting only two film peaks with appreciable intensity and indicating a plane spacing of $2.196~\angstrom$ for planes parallel with the surface of the film.
Aside from the clear Mn$_3$Ge(111) and (222) peaks, there are some small ($<1$\% of the Mn$_3$Ge(111) intensity) peaks present which are made visible by the log scale of Fig.~\ref{fig:TH_TTH_Plot}. These peaks could be due to the presence of either small amounts of other Mn$_3$Ge orientations or small amounts of other Mn-Ge phases. The predominant alignment of the film is consistent with the expected epitaxial relationship of Mn$_3$Ge atop SiC. A cubic system with lattice parameter 0.3803~nm (as obtained by the (111) plane spacing) should have an in-plane, pseudo-hexagonal lattice parameter of $a_h = \sqrt{2}a = 0.5336$~nm~\cite{Bauer:2001aca}. This unit cell dimension is accommodated on the SiC(0001) surface by considering a $\left(\sqrt{3}\times\sqrt{3}\right)R30\degree$-type epitaxy which leads to the expected relationship 
\begin{align}
    [1\bar{1}0]\text{Mn}_3\text{Ge}(111)\|[10\bar{1}0]\text{SiC}(0001).
\end{align}
This expected alignment is shown in Fig.~\ref{fig:Structures}, which also highlights that there should be a compressive strain on the film since its in-plane lattice parameter is 0.80\% larger than that of the SiC $R30\degree$ cell.

It is important to note that the plane-spacing of the cubic Mn$_3$Ge(111), hexagonal Mn$_3$Ge(0001) and tetragonal Mn$_3$Ge(112) families of planes are all very similar and would therefore have very similar diffraction patterns when scanning along the $q_z$ direction. To rule out these other polytypes we conducted further diffraction experiments to gain in-plane information about the film.
\begin{figure}[hbt]
    \centering
	\includegraphics[width=1.0\columnwidth]{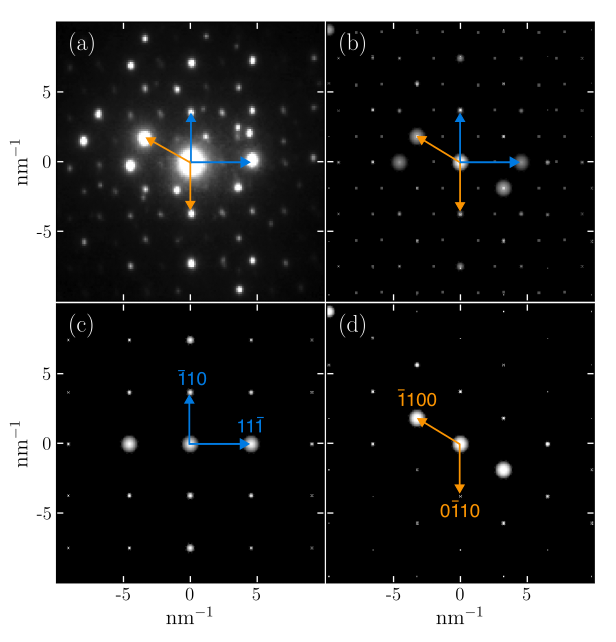}
    \caption{(a) Experimental TEM selected area diffraction pattern and (b) simulated pattern for a 44~nm Mn$_3$Ge layer atop a 44~nm SiC layer with double diffraction spots included, obtained from a sample tilted by 19.5 degrees about the Mn$_3$Ge $[110]$ in-plane direction to align with the Mn$_3$Ge $[112]$ zone-axis. (c) and (d) are simulated patterns from the Mn$_3$Ge and SiC layers, respectively.}
	%\caption{(a) Experimental SADP taken along the Mn$_3$Ge $\left[112\right]$ zone-axis. (b) Simulated TEM diffraction pattern of a 44~nm Mn$_3$Ge layer atop a 44~nm SiC layer along the Mn$_3$Ge $\left[112\right]$ zone-axis with double diffraction effects included. (c) Indexed simulated TEM diffraction pattern of a 44~nm Mn$_3$Ge layer along the Mn$_3$Ge $\left[112\right]$ zone-axis. (d) Indexed simulated TEM diffraction pattern of a 44~nm SiC layer along the Mn$_3$Ge $\left[112\right]$ zone-axis.}
	\label{fig:quad_TEM}
\end{figure}

Transmission electron microscopy (TEM) plan-view samples were prepared by low-angle mechanical polishing and attached to TEM grids with silver epoxy~\cite{Robertson:2006ws}. A selected area diffraction pattern (SADP) was obtained using a 200~kV field-emission microscope after the sample was rotated 19.5$^o$ about the Mn$_3$Ge$\left[1\bar{1}0\right]$ so as to be aligned with the Mn$_3$Ge(112) zone-axis. Along this zone-axis, clear distinction between diffraction patterns owing to the film and to the SiC substrate can be made which support the epitaxial relationship described in Eq.~2. Figure~\ref{fig:quad_TEM} shows the experimental SADP along with a simulated pattern which was obtained using the Bloch-wave method~\cite{Humphreys1979}. Good agreement is observed and peaks can be identified by the individual simulated patterns of Mn$_3$Ge and SiC. Additional peaks that do not appear in the individual Mn$_3$Ge or SiC simulations but do appear in the stacked simulation are due to electrons that have diffracted in both the film and substrate. These double diffraction spots also match well with the experiment.
\begin{figure}[hbt]
    \centering
	\includegraphics[width=1.0\columnwidth]{{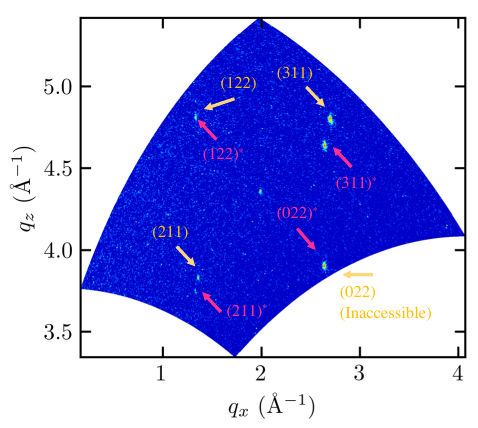}}
	\caption{Reciprocal space map of the $(q_x,q_z)$ plane. Gold, un-starred labels represent diffraction spots from the perfectly cubic system with lattice parameter 0.3803~nm. Magenta, starred labels represent diffraction spots from a cubic system stretched along the [100] direction by 3\%.}
	\label{fig:RSM}
\end{figure}
\begin{figure}[hbt]
  \centering
  \includegraphics[width = 1.0\columnwidth]{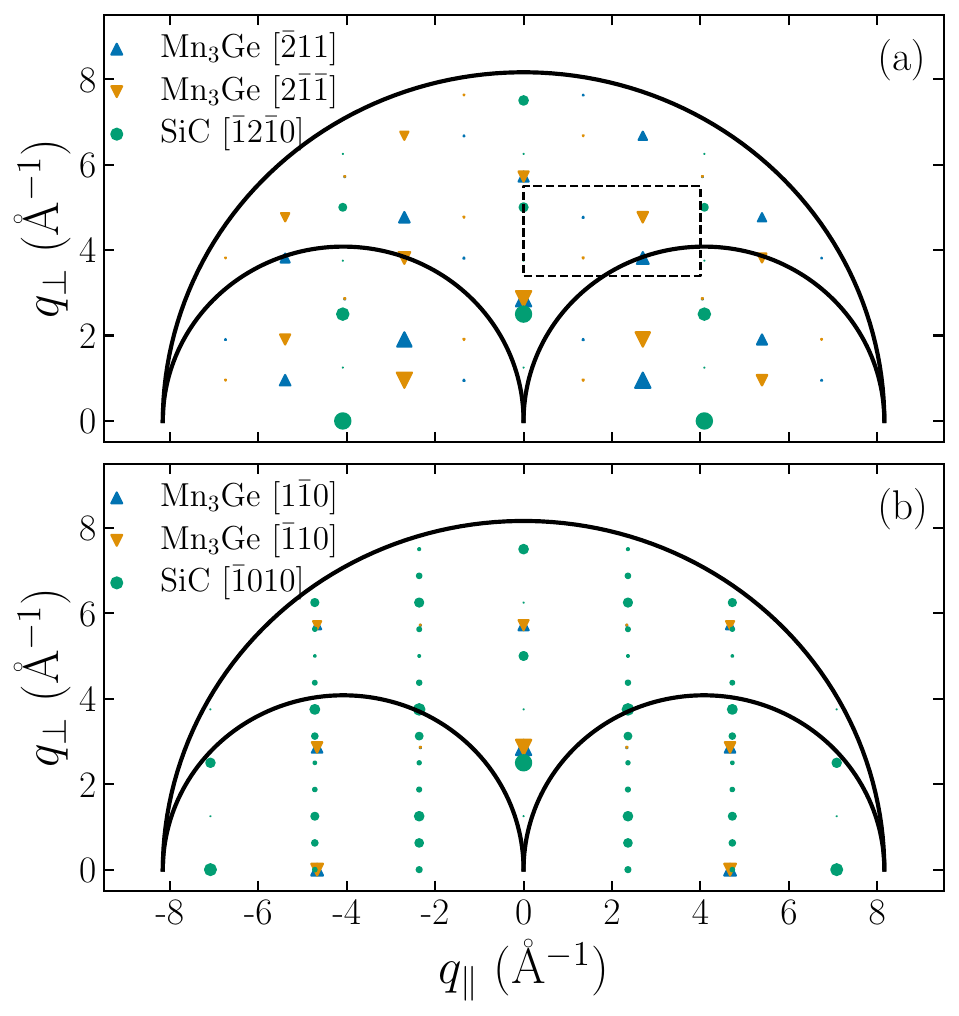}
  \caption{Simulated RSM's with $q_\parallel=q_x\parallel[2\bar{1}\bar{1}]\text{Mn}_3\text{Ge}(111)\|[\bar{1}2\bar{1}0]\text{SiC}(0001)$ (a) and $q_\parallel=q_y\parallel[1\bar{1}0]\text{Mn}_3\text{Ge}(111)\|[1\bar{1}00]\text{SiC}(0001)$ (b). The curved, solid, black lines indicate the area of reciprocal space accessible by a standard diffractometer while the solid black box indicates the region of reciprocal space shown in Fig.~\ref{fig:RSM}.}
  \label{fig:RSM_Sims}
\end{figure}

As further evidence of the cubic structure of our film and the epitaxial relationship given in Eq.~2, we explore the reciprocal space map (RSM) of our films again using Cu K-$\alpha$ X-rays. To distinguish between the two primary, perpendicular, in-plane directions we will denote the direction along SiC$[\bar{1}2\bar{1}0]$ $\vu{x}$ and the direction along SiC$[\bar{1}010]$ $\vu{y}$, with the direction along SiC$[0001]$ reserved for $\vu{z}$. Figure~\ref{fig:RSM} shows a portion of the $(q_x,q_z)$ reciprocal space highlighting the (311) and (211) cubic peaks, which correspond to $[2\bar{1}\bar{1}]\text{Mn}_3\text{Ge}(111)\|[\bar{1}2\bar{1}0]\text{SiC}(0001)$ regions of the film, as well as (022) and (122) peaks, which correspond to $[\bar{2}11]\text{Mn}_3\text{Ge}(111)\|[\bar{1}2\bar{1}0]\text{SiC}(0001)$ regions of the film. The remaining peaks are surmised to be due to a distortion of the perfect cubic structure along the [100] direction. The direction of the distortion is consistent with the [001] direction of a $(112)$-oriented $D0_{22}$ film atop SiC. However, the RSM can only be explained by a \textit{stretching} of the cubic cell along [100], with a resulting c/a ratio of 2.07, which is opposite to the distortion of $D0_{22}$ Mn$_3$Ge with c/a = 1.90. Simulated RSM patterns were prepared with the aid of the open-source software \textit{xrayutilities}~\cite{Kriegner2013} and are shown in Fig.~\ref{fig:RSM} These simulations motivate the portion of reciprocal space probed. In particular, the pattern with $q_x\parallel[2\bar{1}\bar{1}]\text{Mn}_3\text{Ge}(111)\|[\bar{1}2\bar{1}0]\text{SiC}(0001)$ can be used to verify the epitaxial relationship since the film peaks are well separated from the substrate peaks. Conversely, with $q_y\parallel[1\bar{1}0]\text{Mn}_3\text{Ge}(111)\|[1\bar{1}00]\text{SiC}(0001)$, the film peaks are closely aligned with the far higher intensity substrate peaks making this orientation far less useful. 

\subsection{Microscopy}
\begin{figure}[hbt]
    \centering
	\includegraphics[width=1.0\columnwidth]{{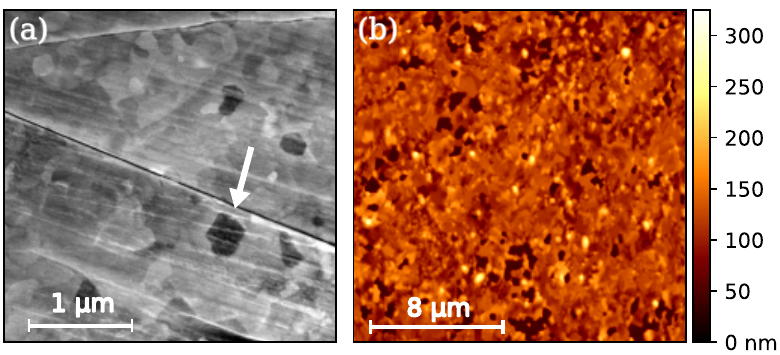}}
	\caption{a) HAADF-STEM image revealing the film island morphology from the white and grey contrast. The arrow points to one of the darker regions where the SiC was exposed. A crack running through the tip of the arrow resulted from the mechanical polishing. b) AFM of the film surface showing the distribution of holes.}
	\label{fig:STEM}
\end{figure}

Scanning transmission electron microscopy (STEM) images were collected with a 200 kV field-emission microscope using the same sample as was mentioned previously. Figure~\ref{fig:STEM}(a) shows the presence of holes in the film: EDS showed no Mn or Ge in these regions.  In the other regions of the sample, EDS found that the film has an excess of Mn with a ratio of Mn:Ge = 3.2(0.2):1.

Surface morphology was investigated with atomic-force-microscopy (AFM). The surface is revealed to be largely continuous with some holes that extend to the substrate (Fig.~\ref{fig:STEM}(b)), as seen in the STEM results. It should be noted that the roughness of these films is very large compared to examples of the hexagonal polytype atop LaAlO$_3$~\cite{Hong:2020prm} and Al$_2$O$_3$/Ru\cite{Hong:2022aplm}. We attribute this roughness to the choice of SiC substrate. Even though the lattice mismatch is extremely small, we expect large differences in surface energies have hindered wetting of the film in a similar manner as was observed in growth of the helimagnet MnSi on SiC~\cite{Meynell2016}. The AFM topographical map gives root-mean-squared roughness of 31.9~nm and a corresponding nominal thickness of 43.9~nm. This thickness is consistent with the peak widths of the (111) and (222) reflections measured in the $\theta$-$2\theta$ X-ray diffraction scan. It is also consistent with the nominal amount of Ge deposited as described by Eq.~\ref{eq:Thickness} if $\alpha=0.8$, which matches the EDS measurement.
%
%\begin{figure}[h]
%    \centering
%	\includegraphics[width=0.65\columnwidth]{{AFM_alone.pdf}}
%	\caption{AFM topographical maps. The topography reveals a largely continuous film with sparse holes which extend downwards to the SiC substrate.}
%	\label{fig:AFM}
%\end{figure}
%

\section{Magnetometry and Transport}
The following magnetometry measurements reveal that the magnetic structure of the cubic films is distinctly different from the other two polytypes. The out-of-plane (OOP) magnetization uncovers a compensation point in the remanent magnetization, indicative of a ferrimagnetic behavior. The corresponding Hall effect measurements provide further insights into the spin structure. We also present in-plane (IP) magnetotransport measurements that show a large anisotropy that derives from the cubic lattice.

Out-of-plane magnetization measurements were performed with a Quantum Design MPMS SQUID magnetometer in the traditional RSO mode with a maximum applied field of 7~T. Transport measurements were performed on a Quantum Design Dynacool PPMS with a maximum applied field of 14T. For transport measurements, samples were cleaved into $\sim(6\times1)$~mm rectangles with constant-current-supplying leads $I^+$, $I^-$, transverse voltage leads $V_{xy}^+$, $V_{xy}^-$ and longitudinal voltage leads $V_{xx}^+$, $V_{xx}^-$ wire-bonded to the surface using 20~$\mu$m aluminum wire. Five constant current leads are used such that the electric field is highly uniform in the region where the voltage leads are attached, as verified by numerical solution of Poisson's equation for this geometry.
\begin{figure}[htb]
    \centering
	\includegraphics[width=1.0\columnwidth]{{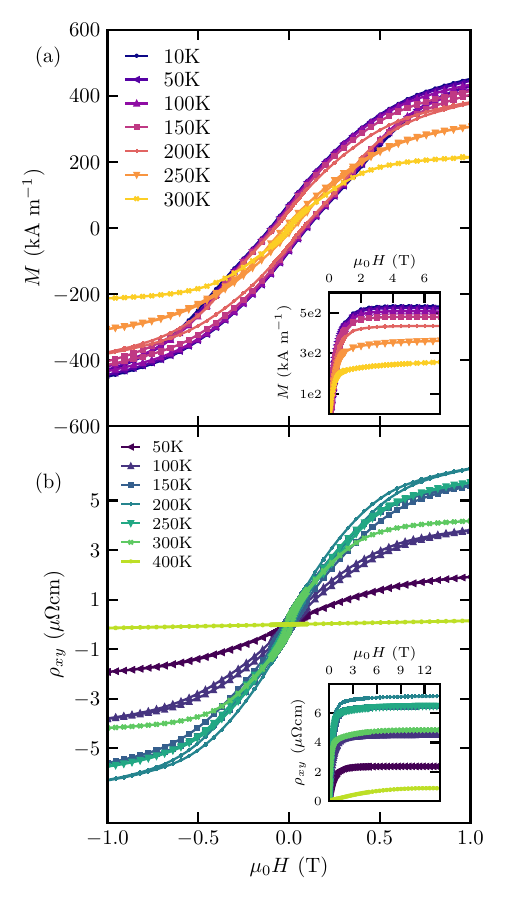}}
	\caption{(a) Magnetization hysteresis loops with external field parallel to SiC$\left[0001\right]$. The diamagnetic susceptibility of the SiC substrate has been subtracted. (b) Anti-symmetrized transverse resistivity hysteresis loops with external field parallel to SiC$\left[0001\right]$. The conventional, linear Hall effect has been subtracted. Insets show the high field behaviors of $M$ and $\rho_{xy}$.}
	\label{fig:M_vs_H_Loops_7T_kAm_SiC_sub}
\end{figure}

\subsection{Out-of-Plane Field}
Figure~\ref{fig:M_vs_H_Loops_7T_kAm_SiC_sub} shows $M$-$H$  and the anti-symmetrized $\rho_{xy}$-$H$ hysteresis loops at various temperatures ranging from 400~K down to 10~K. The linear, diamagnetic susceptibility of the SiC substrates and the linear, conventional Hall effect have been subtracted away from the presented data. It should be noted that although there appears to be saturation in $\rho_{xy}(H)$ for all temperatures considered, there is a small non-linearity that persists up to 14~T which becomes less prevalent as temperature is reduced.

\begin{figure}[htb]
    \centering
	\includegraphics[width=0.85\columnwidth]{{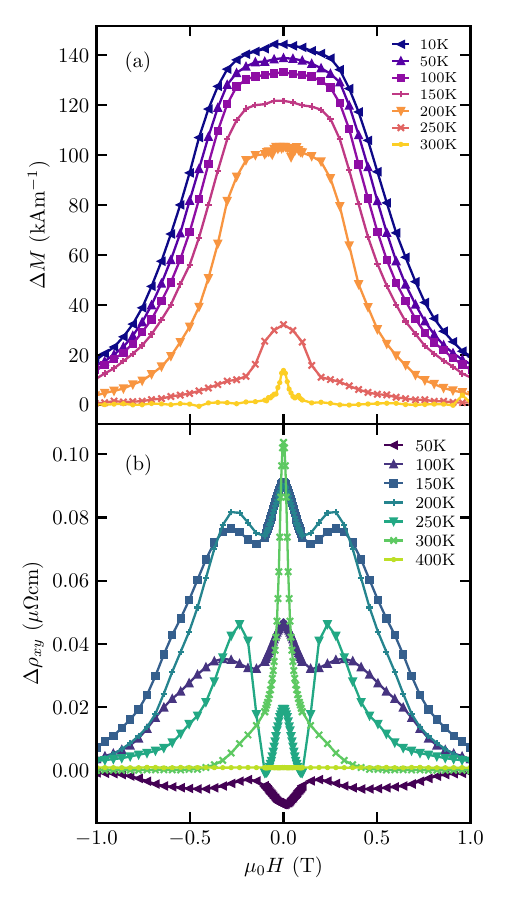}}
	\caption{Differences between the field-decreasing and field-increasing (a) magnetization and (b) transverse resistivity extracted from the isothermal hysteresis loops presented in Fig.~\ref{fig:M_vs_H_Loops_7T_kAm_SiC_sub}.}
	\label{fig:Delta_H_pxy}
\end{figure}

\begin{figure}[htb]
    \centering
	\includegraphics[width=1.0\columnwidth]{{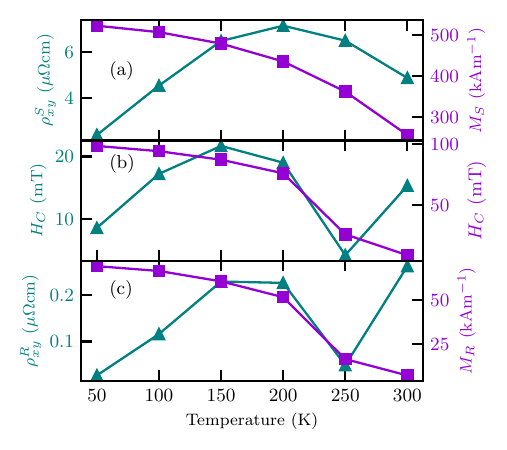}}
\caption{Temperature dependence of (a) saturation magnetization $(M_S)$ and saturation transverse resistivity $\rho_{xy}^S$, (b) coercive fields of $M$ and $\rho_{xy}$ and (c)  magnetization $M_R$ and remanent transverse resistivity $\rho_{xy}^R$. All quantities were extracted from the series of isothemal hysteresis loops shown in Fig.~\ref{fig:M_vs_H_Loops_7T_kAm_SiC_sub}.}
	\label{fig:pxy_M_Compare}
\end{figure}

Figure~\ref{fig:Delta_H_pxy} shows the difference between the field-decreasing branch of the hysteresis loops and the field-increasing branch. Clear deviation between the behavior of $\rho_{xy}(H)$ and $M(H)$ is observed. 
These differences are summarized in Fig.~\ref{fig:pxy_M_Compare}. While the coercive field $H_C$, saturation $M_S$ and remanencence $M_R$ of the magnetization are all monotonically increasing with decreasing temperature, the complementary values of $\rho_{xy}$ show a more complex dependence. Notably, at $\sim250$~K, there is a sharp decrease in the remanent Hall resistivity, $\rho_{xy}^R$, and a change in character of $\Delta\rho_{xy}$.
The significance of this temperature is highlighted in zero-field-warming (ZFW) measurements of the remanent OOP magnetization. Field cooling at 7T(-7T) followed by ZFW shows a ferromagnetic-type behavior up to $T_{C1} = 260$~K, followed by a magnetization reversal above $T_{C1}$. This is contrary to the strictly positive(negative) remanent moments derived from the isothermal hysteresis loops above. We attribute this reversal to an antiferromagnetic coupling between Mn moments on the 3c sites with Mn moments on the Ge 1a sites leading to a ferrimagnetic-like behavior. 
\begin{figure}[htb]
    \centering
	\includegraphics[width=0.85\columnwidth]{{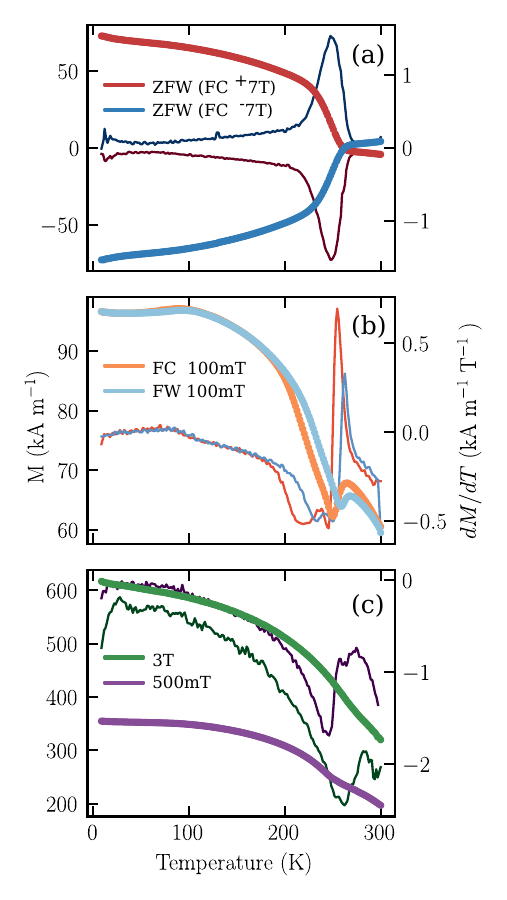}}
\caption{(a) ZFW curves and their derivatives after field cooling in a 7~T and -7~T field. (b) 100~mT FC and FW curves and their derivatives showing clear thermal hysteresis. (c) 500~mT and 3~T FC curves and their derivatives. In all plots, the left axis corresponds to the thick $M(T)$ curves while the right axis corresponds to the thin $dM(T)/dT$ curves.}
	\label{fig:M_vs_T}
\end{figure}
This thermally onset transition is also observed in field-cooled/field-warmed scans at various values of $\mu_0H$, manifesting as a spike in $dM/dT$ and moving to slightly higher temperatures as the external field increases. At low fields, there is substantial thermal hysteresis which recedes as the applied field increases in magnitude.

\begin{figure}[htb]
    \centering
	\includegraphics[width=1.0\columnwidth]{{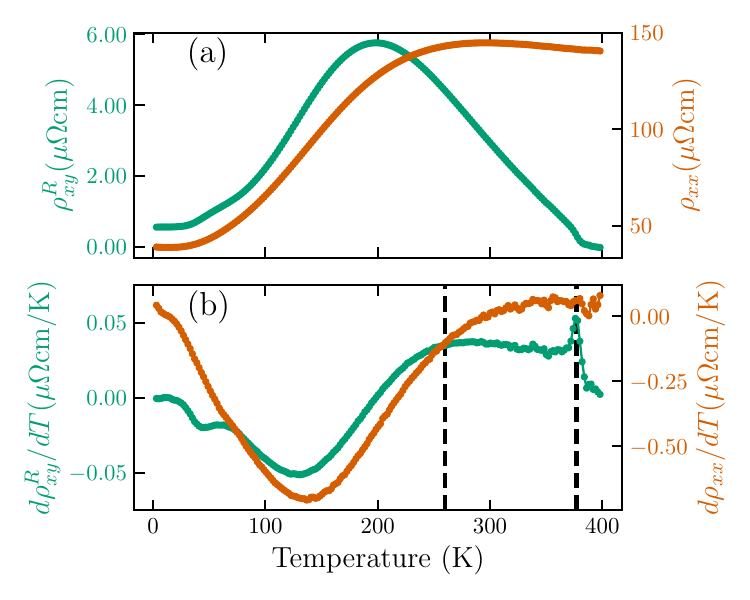}}
\caption{Temperature dependence of (a) $\rho_{xy}^R$ and $\rho_{xx}$ upon warming after field-cooling at 5~T and (b) the derivatives of $\rho_{xy}^R$ and $\rho_{xx}$. Vertical lines have been added at the temperatures $T_{C1}=260$~K and $T_{C2}=377$~K.}
	\label{fig:pxy_temp}
\end{figure}

A more comprehensive investigation of the temperature dependence of $\rho_{xy}$ was achieved by field cooling at 5~T and zero-field warming resulting in a temperature dependence $\rho_{xy}^{0+}(T)$ followed by field cooling at -5~T and zero-field warming resulting in a temperature dependence $\rho_{xy}^{0-}(T)$. By examining $\rho_{xy}^R=\left(\rho_{xy}^{0+}(T) - \rho_{xy}^{0-}(T)\right)/2$, we obtain the remanent transverse resistivity as a function of $T$. This temperature dependence is shown in Fig.~\ref{fig:pxy_temp}, along with the temperature dependence of $\rho_{xx}$ at zero field. 
Possibly due to the higher maximum temperature of this measurement, the remanent $\rho_{xy}^R(T)$ on zero-field-warming exhibits no discernible feature at $T_{C1}$ and is substantially higher than the remanent values obtained from the hysteresis loops in Fig.~\ref{fig:M_vs_H_Loops_7T_kAm_SiC_sub}. Instead, after peaking at 200~K, it linearly decreases until it abruptly approaches zero at $T_{C2} = 377~$K, comparable to the $T=400$~K reported in bulk cubic crystals~\cite{Takizawa:2002jpcm}. The sharp peak in $d\rho_{xy}^R/dT$ at $T_{C2}$ indicates the onset of a remanent magnetization and provides an estimate for the Curie temperature.  This temperature could not be confirmed with the SQUID measurements because of the temperature limit of the sample holder for this instrument, but the presence of a remanent magnetization in SQUID at 300 K places a lower bound for the onset of magnetic order.
%\textbf{Caveat:} It seems likely upon further inspection of this temperature dependence that ``zero-field" was not actually zero-field. Looking at the temperature dependence of the remanent moment as procured from the hysteresis loops, we can see a large discrepancy. In fact, the temperature dependence from the method outlined above can be well-reproduced if we assume that $\rho_{xy}^{0+}(T)$ was taken at 0.1T and $\rho_{xy}^{0-}(T)$ was taken at -0.1T. Vortices in the superconducting magnet are likely responsible, though the temperature history argument made above may have some validity.

\subsection{Field In-Plane}
\begin{figure}[htb]
    \centering
	\includegraphics[width=1.0\columnwidth]{{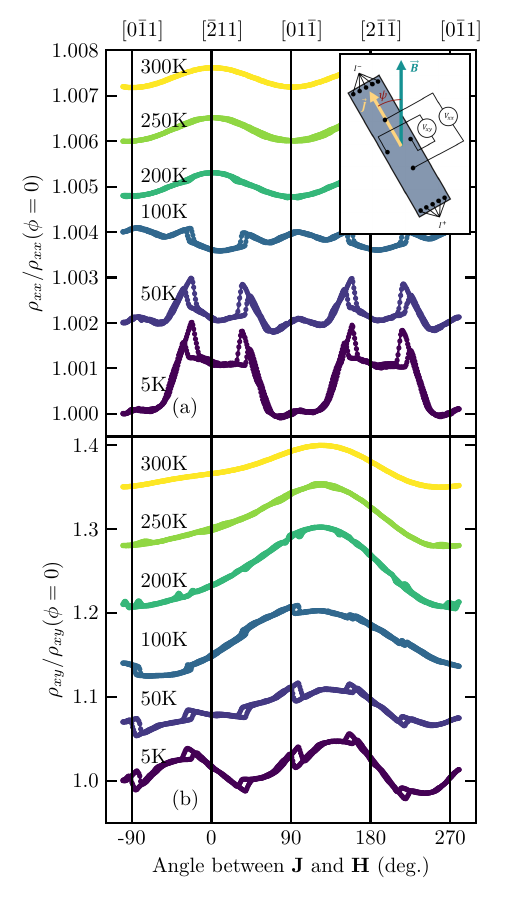}}
	\caption{Constant temperature scans of (a) AMR, $\rho_{xx}(\psi)/\rho_{xx}(\psi=0)$ and (b) PHE, $\rho_{xy}(\psi)/\rho_{xy}(\psi=0)$ at $\mu_0H=3$~T. The inset of (a) shows the experimental geometry.}
	\label{fig:AMR_PHE_vs_T}
\end{figure}

For in-plane measurements, the sample was mounted to a rotating stage in a manner outlined in the inset of Fig.~\ref{fig:AMR_PHE_vs_T}(a). The angle the external field makes with the direction of the current density, $\vb{J}$, and hence the SiC$[2\bar{1}\bar{1}0]$ direction, will be defined as $\psi$. 

\begin{figure}[htb]
    \centering
	\includegraphics[width=1.0\columnwidth]{{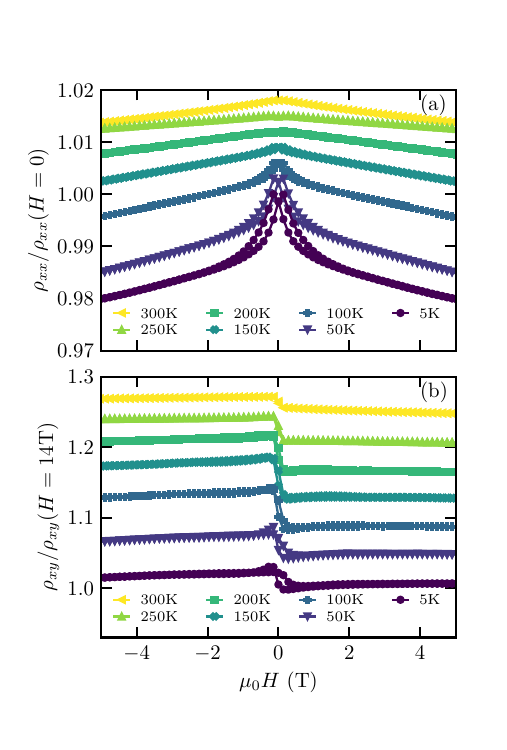}}
	\caption{Constant temperature hysteresis loops of (a) LMR and (b) PHE. These measurements were performed with the applied field parallel to the current density, $\psi=0\degree$. The data have been offset vertically for clarity.}
	\label{fig:AMR_PHE_Loops}
\end{figure}

Isothermal longitudinal magnetoresistance (LMR) and planar Hall effect (PHE) hysteresis loops were conducted at various temperatures, as shown in Fig.~\ref{fig:AMR_PHE_Loops}. These measurements were performed with $\psi=0\degree$, such that the field was along the direction of the current density. Negative LMR is observed for all temperatures, with hysteresis beginning to show at 100~K and growing as the temperature is reduced to 5~K. There is a small, time-reversal-anti-symmetric portion of the PHE with a coercive field of around 0.01~T at 300~K and growing to around 0.08~T at 50~K. This could be attributed to a small misalignment of the film, leading to a component of the OOP Hall effect imparting a contribution.
%but it is worth noting that similar field dependence is observed in the form of a Kerr rotation at 300~K for in-plane field sweeps. 
%

\begin{figure}[htb]
    \centering
	\includegraphics[width=0.85\columnwidth]{{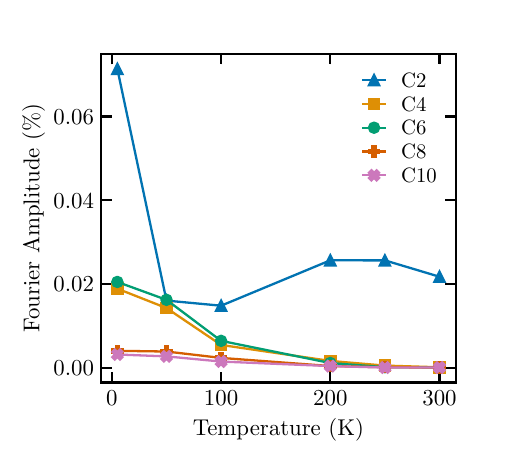}}
	\caption{Relative Fourier components of the average AMR for angle increasing and angle decreasing as a function of temperature at a constant field of $\mu_0H=3$~T.}
	\label{fig:Fourier_T}
\end{figure}

Anisotropic magnetoresistance (AMR) $\left(\rho_{xx}(\psi)\right)$ and PHE $\left(\rho_{xy}(\psi)\right)$ were explored by rotating the angle $\psi$ through $380\degree$ (from -100$\degree$ to 280$\degree$) first at various temperatures with a constant applied field $\mu_0 H=3$~T. At 300~K, typical AMR behavior is observed with $\rho_{xx} \propto \cos^2(\psi)$ and  $\rho_{xy} \propto \sin(2\psi)$. However, upon cooling, clear hysteresis arises, with discontinuous jumps occurring in both $\rho_{xx}$ and $\rho_{xy}$ at intervals of $\sim60\degree$. This change in behavior from high to low temperature is captured with Fourier analysis of the AMR/PHE as shown in Fig.~\ref{fig:Fourier_T}. At higher temperatures, the conventional, 2-fold rotational symmetry is dominant, but as temperature falls, other even Fourier components contribute, with 6-fold and 4-fold components the primary among them. 

\begin{figure}[htb]
    \centering
	\includegraphics[width=1.0\columnwidth]{{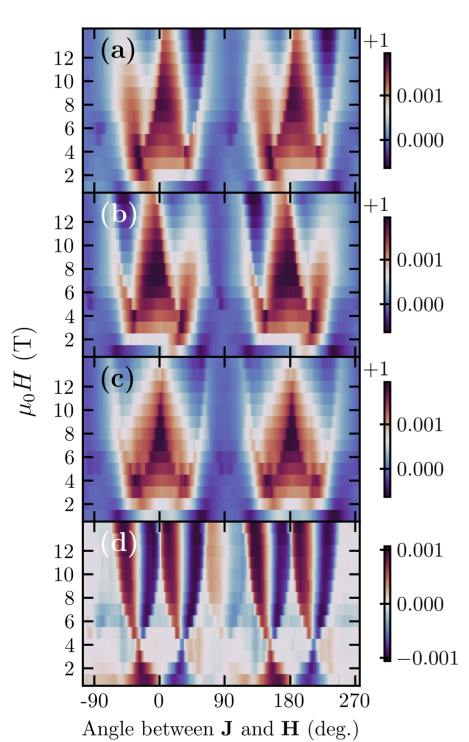}}
	\caption{AMR, $\rho_{xx}(\psi)/\rho_{xx}(\psi=0)$ for (a) $\psi$ increasing, (b) $\psi$ decreasing, (c) the average of increasing and decreasing and (d) the difference between increasing and decreasing.}
	\label{fig:AMR_vs_H}
\end{figure}

\begin{figure}[htb]
    \centering
	\includegraphics[width=1.0\columnwidth]{{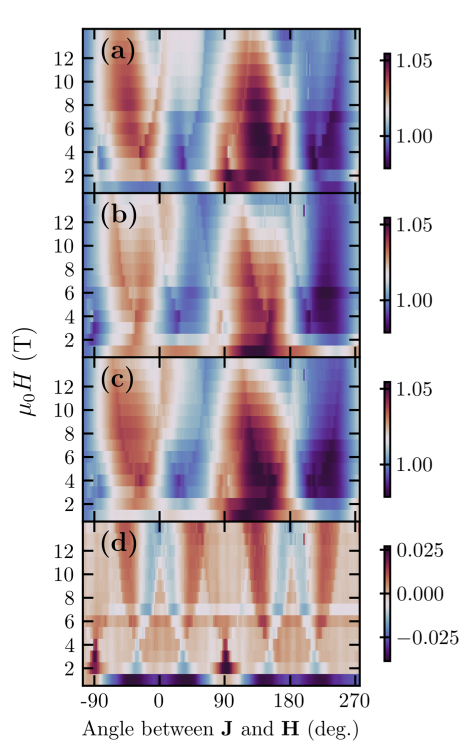}}
	\caption{PHE, $\rho_{xy}(\psi)/\rho_{xy}(\psi=0)$ for (a) $\psi$ increasing, (b) $\psi$ decreasing, (c) the average of increasing and decreasing, and (d) the difference between increasing and decreasing.}
	\label{fig:PHE_vs_H}
\end{figure}

\begin{figure}[htb]
    \centering
	\includegraphics[width=0.85\columnwidth]{{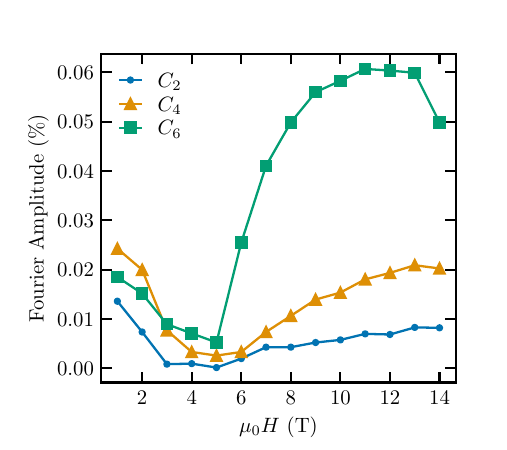}}
	\caption{Relative Fourier component amplitudes of the difference in the angle increasing AMR and angle decreasing AMR as a function of the applied field at 5~K.}
	\label{fig:AMR_Fourier}
\end{figure}

To further examine the low temperature, unconventional AMR/PHE, rotation measurements were performed with varying applied fields at a constant temperature of 5~K. These are shown in Fig.~\ref{fig:AMR_vs_H} and Fig.~\ref{fig:PHE_vs_H}. Clear hysteresis is present for all fields considered, with the discontinuous jumps present at $60\degree$ intervals appearing to be amplified and smeared out into more continuous transitions as the field is increased. The difference between the angle increasing and angle decreasing AMR and PHE shows a very clear 6-fold rotational symmetry, as is captured by the Fourier analysis shown in Fig.~\ref{fig:AMR_Fourier}.
\section{Density Functional Theory Calculations}
\subsection{Computational Details}
The density functional theory (DFT) simulations were carried out with the Vienna Ab~initio Simulation Package (VASP) \cite{Kresse:1996prb2} within the generalized gradient approximation (GGA) \cite{Perdew:1996prl} and the projector augmented wave (PAW) method \cite{Kresse:1999prb}. The calculations included a non-collinear treatment of the magnetization, spin-orbit coupling, and the removal of symmetry recognition. The following simulation parameters were adopted: kinetic energy cut-off of 700~eV, Monkhorst-Pack \cite{Monkhost:1976prb} Brillouin zone {\bf k}-sampling of 10$\times$10$\times$10 for the cubic phase and 8$\times$8$\times$8 for the hexagonal phase, Methfessel-Paxton smearing \cite{Methfessel:1989prb} of 0.15~eV, electronic energy convergence of 10$^{-8}$~eV and relaxation energy convergence of 10$^{-6}$~eV. Ground state lattice constants were obtained by relaxing the atomic positions and cell size/shape. \\

\subsection{Results}
We performed DFT simulations on three ordered phases of Mn$_3$Ge: the $D0_{19}$ structure with AFM order and the cubic $L1_2$ phase with both AFM and FM orders. For the cubic AFM phase, we considered the $T_{1g}$ chiral AFM structure shown in Fig.~\ref{fig:Structures}(e) that is reported in the other $L1_2$ cubic AFMs~\cite{Kren:1966pl, Tomeno:1999jap, Sakuma:2002prb, Ikeda:2003jpsj}. We found that the local magnetic moment of Mn is 2.446\,$\mu_B$ with an average angle of 84.36$^\circ$ relative to the [111] direction (small canting out of the (111) plane), as shown in Table~\ref{tab:DFT}. The angles of the Mn moments are found to vary slightly between 84.30$\degree$ and 84.40$\degree$. This out-of-plane tilting leads to a total magnetic moment of 0.76~$\mu_B$ aligned 0.04$\degree$ relative to the [111] direction. A strained cubic lattice, converted into a hexagonal cell, was also tested based on experimental estimates of the lattice constants $a$\,=\,5.390\,\AA\, and $c$\,=\,6.285\,\AA. The calculated local magnetic moments and their average orientation relative to the $c$-axis (equivalent to the [111] direction in the cubic cell) are 2.565\,$\mu_B$/Mn and 88.51$^\circ$, which represents an increase of $\approx$\,5\%. There is also an energy increase of 17\,meV/atom with the strained cell. \\

For the hexagonal phase, we used the $E_{1g}$ antichiral spin structure determined by spherical neutron polarimetry, Fig.~\ref{fig:Structures}(b) \cite{Soh:2020prb}. Our DFT-calculated lattice parameters for the hexagonal AFM phase are within 2\% of the experimental values of $a$\,=\,5.339-5.347\,\AA\,and $c$\,=\,4.314\,\AA\, \cite{Yamada:1988pb} . The hexagonal AFM structure exhibits a measured local magnetic moment of 2.65(2)\,$\mu_B$/Mn \cite{Soh:2020prb}, which is higher than our value of 2.291\,$\mu_B$/Mn. In another DFT study \cite{Yang:2017njp}, a theoretical value of 2.7\,$\mu_B$/Mn was obtained using the experimental lattice constants and the local density approximation for exchange-correlation. When adopting the same lattice parameters, we obtain a magnetic moment of 2.569\,$\mu_B$/Mn, in closer agreement with the experimental value.\\

\begin{figure}[hbt]
  \centering
  \includegraphics[width = 1.0\columnwidth]{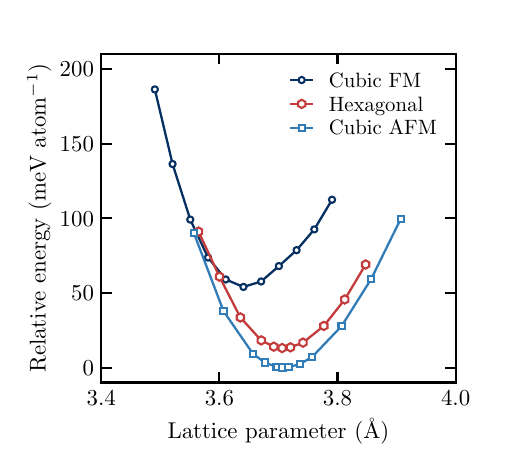}
  \caption{The relative energy per atom of the three structures considered in the DFT simulations.}
  \label{fig:DFT}
\end{figure}

The spin-configuration used for the cubic phase with a FM order is shown in Fig.~\ref{fig:Structures}(c). The DFT-computed lattice constant is $\approx$\,4\% smaller than the value reported in bulk samples, 3.81\,\AA, and our magnetic moment is not in good agreement with the value of 0.87$\mu_B$/Mn \cite{Takizawa:2002jpcm}.  It should be noted that Ge exhibits a local magnetic moment that is aligned opposite to that of Mn, suggesting that this phase is ferrimagnetic. \\

Energetically, the cubic AFM phase is the lowest energy configuration, as shown in Fig.~\ref{fig:DFT}, followed by hexagonal AFM with an additional energy of 13\,meV/atom and cubic FM with an additional energy of 54\,meV/atom. Finally, we note that when the cell shape was allowed to change, each phase showed a tiny structural distortion leading to a negligible change in the properties of Mn$_3$Ge. \\

%\begin{table*}[hbtp]
%\begin{tabular}{| c || c | c | c | c | c | c | }
%\hline
%                          &  	Lattice constant  	& 	Local magnetic  			& 	Angle of local             		& 	Total magnetic 		& 	Angle of total 				& 	Energy/atom  \\
%                          & 	(\AA)             	&    moment ($\mu_B$)      	&    magnetic moment         	&    	moment ($\mu_B$)      &    magnetic moment			&   	(eV)             \\
%\hline\hline
%Cubic AFM         	& 	3.707                  	&   	2.446, 0.010 				&  	84.36$^\circ$, 180$^\circ$ to [111]    	&   	0.760                  		&    0.04$^\circ$ to [111]  		& 	-8.040965  \\
%Hexagonal AFM 	& 	5.241, 4.229         &    2.291, 0.000          		&   	90.00$^\circ$, --  to [001]  				&   	0.014                    	&  	89.68$^\circ$  to [001]  	& 	-8.027891 \\
%Cubic FM           	& 	3.641                 	&  	1.691, 0.116 				&  	0$^\circ$, 180$^\circ$  to [100]  		&  	5.216                        	&  	0$^\circ$  to [100]   		&  	-7.986951
%\\
%\hline
%\end{tabular}
%\caption{DFT-computed properties of Mn$_3$Ge. For the cubic AFM phase, the angle of the local moment represents the average with values ranging between 84.30-84.40$^{\circ}$. For the hexagonal phase, the two numbers correspond to the $a$ and $c$ lattice parameters. For the local magnetic moment, the first and second number correspond to the Mn and Ge components, respectively.}
%\label{tab:DFT}

%\end{table*}

\begin{table*}[hbtp]
\begin{ruledtabular}
\begin{tabular}{c | cccccc }
%\hline
                          &  	Lattice constant  	& 	Local magnetic  			& 	Angle of local             		& 	Total magnetic 		& 	Angle of total 				& 	Energy/atom  \\
                          & 	(\AA)             	&    moment ($\mu_B$)      	&    magnetic moment         	&    	moment ($\mu_B$)      &    magnetic moment			&   	(eV)             \\
\hline
Cubic AFM         	& 	3.707                  	&   	2.446, 0.010 				&  	84.36$^\circ$, 180$^\circ$ to [111]    	&   	0.760                  		&    0.04$^\circ$ to [111]  		& 	-8.040965  \\[4pt]
Hexagonal AFM 	& 	5.241, 4.229         &    2.291, 0.000          		&   	90.00$^\circ$, --  to [001]  				&   	0.014                    	&  	89.68$^\circ$  to [001]  	& 	-8.027891 \\[4pt]
Cubic FM           	& 	3.641                 	&  	1.691, 0.116 				&  	0$^\circ$, 180$^\circ$  to [100]  		&  	5.216                        	&  	0$^\circ$  to [100]   		&  	-7.986951
\\
%\hline
\end{tabular}
\end{ruledtabular}
\caption{DFT-computed properties of Mn$_3$Ge. For the cubic AFM phase, the angle of the local moment represents the average with values ranging between 84.30-84.40$^{\circ}$. For the hexagonal phase, the two numbers correspond to the $a$ and $c$ lattice parameters. For the local magnetic moment, the first and second number correspond to the Mn and Ge components, respectively.}
\label{tab:DFT}

\end{table*}

To compare to the momentum-space Berry curvature resulting in the AHE in the cubic Mn$_3$Ir and Mn$_3$Pt, the open-source \textit{Wannier90}~\cite{Mostofi2014} was used to calculate the anomalous Hall vector of the cubic AFM spin-configuration described previously. For the Berry curvature calculation, semicore states were included in the Wannierization process. This resulted in 72 maximally localized Wannier functions. A uniform, $200\times200\times200$ $k$-point mesh was used for the Brillouin zone integration of the static anomalous Hall conductivity. The calculation indicates the presence of an intrinsic anomalous Hall vector with a magnitude of 155.7~S/cm at the Fermi level of the stoichiometric system. The Hall vector is directed along the $\left[111\right]$ crystallographic direction, consistent with calculations completed for Mn$_3$Ir and Mn$_3$Pt.

\section{Discussion}
While the $M(H)$ loops in Fig.~\ref{fig:M_vs_H_Loops_7T_kAm_SiC_sub}(a) are qualitatively similar to those of the bulk crystals reported in Ref.~\onlinecite{Takizawa:2002jpcm}, with comparable values of $M_S$, our other results do not support the claim that the Mn$_{3.2}$Ge films exhibit simple ferromagnetism. The average moment of 0.98~$\mu_B$/Mn is only 58\% of the DFT calculated value. The reversal of the remanent moment at a compensation point $T = 260$~K indicates that there are at least two different magnetic sites in the crystal that are coupled antiferromagnetically. The topological contribution to the Hall signal, evident from the different hysteretic behavior of $M(H)$ and $\rho_{xy}$ in Fig.~\ref{fig:Delta_H_pxy}, indicates that these spins form a non-collinear configuration.  Furthermore, the angular dependence of the PHE in Fig.~\ref{fig:AMR_vs_H} shows hysteresis up to at least 14~T, well above the apparent saturation field in the magnetometry measurements.  This may be related to a competition between the magnetocrystalline anisotropy and the large exchange field between antiferromagnetically coupled spins.  However, the results of these measurements are difficult to interpret without micromagnetic modelling and more detailed information about the atomic spin structure.

In addition to scalar spin chirality, there are likely significant contributions to the Hall effect from band structure topology. The low temperature longitudinal conductivity $\sigma_{xx} = 3\times 10^4$~S/cm, places this material in the the so-called \emph{good metal} regime, where intrinsic contributions to the AHE dominate~\cite{Miyasato:2007prl}.  It is interesting to note that the size of the Hall conductivity $\sigma_{xy} = 1587$~S/cm at $T=50$~K is the highest reported for kagome magnets~\cite{Chen:2021nc}: it is larger than the kagome ferromagnet Co$_3$Sn$_2$S$_2$, ($\sigma_{xy} = 1130$~S/cm~\cite{Liu:2018np}) and the frustrated kagome ferromagnet Fe$_3$Sn$_2$, ($\sigma_{xy} = 402$~S/cm~\cite{Wang:2016prb}).   

To understand the non-collinear structure it is important to consider the influence of chemical disorder.  This disorder could be caused by the excess Mn in our films. The other report on cubic Mn$_3$Ge does not give any measure of the composition of the samples for us to compare to~\cite{Takizawa:2002jpcm}. To find a point of reference, we turn to the $D0_{19}$ compounds.  Excess Mn is commonly used to help stabilize the Mn$_3$Ge hexagonal phase, but for low enough concentrations, this does not destroy the chiral antiferromagnetic structure: the 120$\degree$ inverse triangular structure was found for $D0_{19}$ Mn$_{3.1}$Ge~\cite{Kadar:1971ijm}  and Mn$_{3.34}$Sn~\cite{Zimmer:1973cp}. At higher concentrations, glassy magnetic behavior can appear, as found for $D0_{19}$ Mn$_{3.44}$Sn~\cite{Feng:2006prb}, which shows a bifurcation between the zero-field cooled and field-cooled $M$-vs-$T$ data.  This occurs in samples with residual resistivity ratios lower than 4~\cite{Ikhlas:2020jpsj}, which is comparable to our films.

Insight into the impact of disorder on the magnetic structure can be gained by comparing the $L1_2$ and $D0_{22}$ Mn$_3$Ge polytypes. In the case of disorder in a stoichiometric $L1_2$ phase, an exchange of Mn and Ge between the $1a$ and $3c$ sites drives the material towards the $D0_{22}$ phase. The Mn anti-site defects would be expected then to couple antiferromagnetically to the $3c$ Mn. The change would try to drive a distortion of the structure along the [100] direction.  Surprisingly from Fig.~\ref{fig:RSM}, we see a small elongation along the [100] direction, rather that the contraction expected from the $D0_{22}$ structure.  However, the tetragonal phase is not as well matched to SiC(0001), which may be the reason that the system remains nearly cubic. The fact that the moment of the $D0_{22}$ phase is far less than what we observe is further evidence that this phase is not formed.  Based on DFT, the Mn on the $3c$ sites would like to form a 120$\degree$ chiral structure. However, the expected antiferromagnetic interaction between the Mn$_{3c}$ and Mn$_{1a}$ would cause a large canting of the Mn away from the (111) plane.

Future studies of the effects of excess Mn should be of great interest.  Given that the DFT simulations predict that the stoichiometric compound should have a chiral AFM structure, the spin structure should be quite sensitive to chemical tuning.  The material potentially offers the opportunity of creating frustration-stabilized skyrmions~\cite{Leonov:2015nc}, or other textures such as those found in Fe$_3$Sn$_2$~\cite{Hou:2017am, Du:2022prl}.

\section{Summary}
We have successfully synthesized Mn$_{3.2}$Ge films with a cubic kagome structure. Our magnetometry and magnetotransport measurements indicate that this is an interesting topological material worthy of future exploration.  The giant anomalous Hall conductivity is the largest reported so far for the kagome magnets, possibly due to topological features in the underlying kagome crystal structure.  A non-coplanar spin structure is also likely contributing a real-space Berry curvature component to the AHE signal. Further improvements in film morphology could make this an attractive material for device applications given its above-room-temperature Curie temperature and giant Hall response.

\section{Acknowledgments}
We acknowledge financial support from the Natural Science and Engineering Research Council (Canada).  We would like to thank M.~L.~Plumer, B.~W.~Southern and A.~Zelenskiy for helpful discussions.  We thank Laurent Kreplak for assistance with AFM measurements and Christian Lupien for assistance with transport measurements. Computational resources were provided by the Digital Research Alliance of Canada.

\clearpage
\appendix
\section{DFT-computed atomic structure and magnetic moments}
\begin{table}[htb]

\begin{ruledtabular}
\begin{tabular}{c | cc  }
           	& 	Atomic position 				& 	Local magnetic moment ($\mu_B$) 	\\
\hline
Mn         &   	(0.16133 0.32266 0.75000)         &    (0 2.291 0)              				\\
Mn         &    (0.83860 0.67734 0.25000)         &    (0 2.291 0)               			\\
Mn         &    (0.32266 0.16133 0.25000)         &   	(1.982 -1.150 0)                      		\\
Mn         &    (0.83860  0.16133 0.25000)        &    (-1.983 -1.147 0)          			\\
Mn         &    (0.67734 0.83867 0.75000)         &    (1.982 -1.150 0)             			\\
Mn         &    (0.16133 0.83867 0.75000)         &    (-1.983 -1.147 0)                		\\
Ge         &    (0.33333 0.66666 0.25000)         &    (0 0 0)                   				\\
Ge         &    (0.66666 0.33333 0.75000)         &    (0 0 0)             					\\
\end{tabular}
\end{ruledtabular}

\caption{Hexagonal AFM Mn$_3$Ge. The lattice vectors are $a$$\cdot$[1 0 0], $a$$\cdot$[-1/2 $\sqrt{3}$/2 0], $c$$\cdot$[0 0 1], where $a$\,=\,5.241\,\AA\, and $c$\,=\,4.229\,\AA. The total magnetic moment vector is [-0.0035 -0.0135 0.0000]\,$\mu_B$. The atomic positions are expressed in terms of the lattice vectors.}
\label{tab:DFT_hexa_AFM} 

\end{table}
\begin{table}[htb]
\begin{ruledtabular}
\begin{tabular}{c | c c   }
       	&  	Atomic position  		& 	Local magnetic moment ($\mu_B$) 	\\
\hline
Mn        	&   	(0.0 0.5 0.5)           	&    (-1.849 1.132 1.133)               		\\
Mn        	& 	(0.5 0.0 0.5)             	&    (1.132 -1.850 1.132)                		\\
Mn        	&  	(0.5 0.5 0.0)               &  	(1.134 1.134 -1.847)         			\\
Ge     	&    (0.0 0.0 0.0)              	&  	(-0.006 -0.006 -0.006)          		\\
\end{tabular}
\end{ruledtabular}

\caption{Cubic AFM Mn$_3$Ge. The lattice vectors are $a$$\cdot$[1 0 0], $a$$\cdot$[0 1 0], $a$$\cdot$[0 0 1], where $a$\,=\,3.707\,\AA. The total magnetic moment vector is [0.4387 0.4386 0.4394] $\mu_B$. The atomic positions are expressed in terms of the lattice vectors.}
\label{tab:DFT_cubic_AFM} 

\end{table}
\begin{table}[htb]

\begin{ruledtabular}
\begin{tabular}{ c | c  c   }
        	&  	Atomic position  	& 	Local magnetic moment ($\mu_B$) 	\\
\hline\hline
Mn         &    (0.0 0.5 0.5)         &    (1.690 0 0)               				\\
Mn         & 	(0.5 0.0 0.5)         &    (1.692 0 0)                 			\\
Mn         &  	(0.5 0.5 0.0)         &  	(1.692 0 0)            				\\
Ge         	&   	(0.0 0.0 0.0)        	&    (-0.116 0 0)         					\\
\end{tabular}
\end{ruledtabular}

\caption{Cubic FM Mn$_3$Ge. The lattice vectors are $a$$\cdot$[1 0 0], $a$$\cdot$[0 1 0], $a$$\cdot$[0 0 1], where $a$\,=\,3.641\,\AA. The total magnetic moment vector is [5.2159 0.0004 -0.0002]. The atomic positions are expressed in terms of the lattice vectors.}
\label{tab:cubic_FM} 
\end{table}

\clearpage

\bibliographystyle{apsrev}
%\bibliography{/Users/tedmonchesky/Library/CloudStorage/OneDrive-DalhousieUniversity/bib/primary.bib}

\end{document}